\documentclass[superscriptaddress,aps,pra,twocolumn,showpacs,nofootinbib,longbibliography]{revtex4-1}
\usepackage{amsmath}
\usepackage{latexsym}
\usepackage{amssymb}
\usepackage[colorlinks=true,citecolor=blue,urlcolor=blue]{hyperref}
\usepackage{color}
\usepackage{graphics,epstopdf}
\usepackage{soul}

\begin{document}

\title{Sharing of Non-Local Advantage of Quantum Coherence  by sequential observers}

\author{Shounak Datta}
\email{shounak.datta@bose.res.in}
\affiliation{S. N. Bose National Centre for Basic Sciences, Block JD, Sector III, Salt Lake, Kolkata 700 098, India}

\author{A. S. Majumdar}
\email{archan@bose.res.in}
\affiliation{S. N. Bose National Centre for Basic Sciences, Block JD, Sector III, Salt Lake, Kolkata 700 098, India}

%\author{}
%\email{}
%\affiliation{}

%\author{}
%\email{}
%\affiliation{}

\date{\today}

\begin{abstract}
Non-local Advantage of Quantum Coherence(NAQC) or steerability of local quantum coherence is a strong non-local resource based on coherence complementarity relations.  In this work, we provide an upper bound on the number of observers who can independently steer the coherence of the observer in the other wing in a scenario where half of an entangled pair of spin-$\frac{1}{2}$ particles is shared between a single observer (Bob) in one wing and several observers (Alices) on the other, who can act sequentially and independently of each other. We consider one-parameter dichotomic POVMs for the Alices and mutually unbiased basis in which Bob measures coherence in case of the maximally entangled bipartite qubit state. We show that not more than two Alices can exhibit NAQC when $l_1$-norm of coherence measure is probed, whereas for two other measures of coherence, only one Alice can reveal NAQC within the same framework.
\end{abstract}

\pacs{03.67.-a, 03.67.Mn}

\maketitle

\section{Introduction}

The identifying feature of quantum correlation is that no classical description is adequate for its realization. Following up on the famous Einstein-Podolsky-Rosen(EPR) paradox\citep{EPR}, Schr\"{o}dinger\citep{Sch,Sch1} established the possibility of steering of one subsystem by the other for entangled states. Subsequently, Bell showed that\citep{Bell} a local realistic description based on the classical world view is not consistent with the ideas of Quantum Mechanics (QM). Bell non-locality is a key feature of QM. Many years later, Reid\citep{Reid} and subsequently Wiseman et al\citep{Wiseman,Jones,Cavalcanti} provided an operational formalism of EPR  steering. According to the hierarchy of the strength of correlations for bipartite quantum states,  steerable states form a strict subset of entangled states and a strict superset of Bell non-local states.

The task of steering lies in remotely preparing an ensemble of states for Bob by measuring on the subsystem of Alice while Alice and Bob share a bipartite entangled state, $\sigma_{AB}$, ruling out a  local hidden state (LHS) model for Bob. As quantum coherence is an inherent property of a quantum state, so one may ask whether steering of a quantum state implies steering of quantum coherence and vice-versa. This question was addressed recently in Ref.\citep{Mondal} to establish the criterion of steerability of quantum coherence through constructing the coherence complementarity relations for the local state of Bob under the framework of an LHS model. Several studies on quantum coherence have provided basis-dependent quantifications of it which include $l_1$-norm of quantum coherence\citep{Baumgratz}, relative entropy of quantum coherence\citep{Baumgratz} and skew-information\citep{Girolami,Luo}. Using these three measures of coherence, it has been shown~\citep{Mondal} that steering of a quantum state and its coherence are different intrinsically though steering of quantum coherence is a sufficient condition for steering of quantum state. This non local feature was termed as Non-local Advantage of Quantum Coherence(NAQC).

The connection of quantum coherence with correlations in bipartite systems has been further explored in \citep{Hu}, where it was shown that, NAQC geometrically follows the hierarchy of  quantum correlations as in Bell non-locality, quantum steering and quantum entanglement. The states which manifest NAQC form a subset of Bell non-local correlations and thereby NAQC is proved to be the strongest among all such correlations for all bipartite quantum states. Recently, non-local advantage of quantum coherence has been probed for higher dimensional quantum states, and the interrelation between coherence and steerability is explored further\citep{Hu1}. The complementarity between various coherence steering criteria has also been proposed recently\citep{Mondal1}. Such studies have acquired importance with the development of   quantum coherence  as a resource\citep{Streltsov} in quantum information theory. 

Quantum correlations satisfying no-signalling conditions are generally monogamous, and relaxation of no-signalling implies the violation of monogamy relations. In particular, no-signalling cannot be enforced in scenarios such as the case when half of an entangled pair of two particles is shared by one observer and another half is shared among several observers who measure their particle sequentially and independently of each other\citep{Silva}. It has been shown in\citep{Shiladitya} that if a maximally entangled state is shared between a Bob at one end and several Alices at the other,  a maximum of two Alices can exhibit QM violation of the Bell-CHSH inequality\citep{CHSH}, a result that was conjectured numerically earlier~\citep{Silva}, and later on also demonstrated experimentally~\citep{Expt1,Expt2}. In case of quantum steering, it has been recently shown~\citep{Das} that at most two Alices can demonstrate quantum steering in the same scenario using the CHSH-analogous steering inequality which is a necessary and sufficient criterion for steering\citep{Analog} in such cases. Further, using an $n$-measurement linear steering criterion\citep{Cavalcanti} in  the above scenario, it has been conjectured that optimally $n$ number of Alices can perform quantum steering of Bob's state~\citep{Das}.

In the present work we investigate the issue of sharing of NAQC which is a quantum correlation stronger than Bell nonlocality~\citep{Hu}, by multiple observers on one wing. In light of the above studies we are motivated by the question as to whether the number of the observers who share quantum correlations sequentially depends on the strength of the correlation present.  Specifically, we consider the scenario where multiple Alices share half of an entangled pair of qubits and perform self-contained unbiased unsharp measurements sequentially, whereas a single Bob measures quantum coherence in one of the mutually unbiased basis at the other end. This scenario is compatible with the original description of the LHS model employed in~\citep{Mondal}. NAQC depending on various measures of quantum coherence are different by construction. In case of the $l_1$-norm of NAQC, we find that at most two Alices can steer the coherence of Bob, whereas in the cases of relative entropy of NAQC and skew information of NAQC, we find that one Alice is maximally sufficient for the manifestation of NAQC in the same scenario. 

We arrange this paper in the following way. In Section \ref{2}, we provide a brief description of Non-local Advantage of Quantum Coherence. In Section \ref{3}, we describe the framework of unsharp measurement which is applied by several observers at one side of the bipartite state. In Section \ref{4}, we analyze our scenario and obtain bounds on the number of observers in cases where three different forms of $3$-settings NAQC are probed. We conclude with some summarizing remarks in Section \ref{5}.

\section{Differnet forms of Non-local Advantage of Quantum Coherence} \label{2}

The non-existence of local hidden state(LHS) model demonstrates steering. If Bob holds an ensemble $\lbrace P(\lambda), \rho_B^Q (\lambda)\rbrace$, i.e., every LHS $\rho_B^Q (\lambda)$ depending on hidden variable $\lambda$ occurs with probability $P(\lambda)$, then conditional state on Bob's side contingent upon measuring $A$ with outcome $a$ at Alice's side can be written as\citep{Wiseman},
\begin{equation}
\rho_A^a = \sum_{\lambda} P(\lambda) P(a|A,\lambda) \rho_B^Q (\lambda)
\end{equation}
where, $\sum_{\lambda} P(\lambda)=1$. If such a map exists, then using $P(a_A,b_B)=\operatorname{Tr}[\Pi_B^b \rho_A^a]$, one can write the joint probability function of getting outcome $a$ by doing measurement $A$ at Alice's side and getting outcome $b$ by measuring $B$ at Bob's side as,
\begin{equation}
P(a_A,b_B)=\sum_{\lambda} P(\lambda) P(a|A,\lambda) P_Q(b|B,\lambda)
\label{str}
\end{equation}
where, $P_Q(b|B,\lambda)$ is derived from $\rho_B^Q (\lambda)$ using Born's rule. If the equality of Eq.(\ref{str}) no longer holds, then the state shared between Alice and Bob can be called as steerable. Various steering conditions for bipartite systems have been constructed\citep{Cavalcanti,Analog,Entropic,Fgs,Saunders}. For bipartite qubits, there are a number of quantifiers available in the literature\citep{sw,ros,sf,sc} to measure quantum steering. Similarly, there exist steering frameworks for continuous variable systems also\citep{Reid,Cont1,Cont2,Cont3,Sum}.

To demonstrate an LHS model for Bob, several uncertainty relations, e.g. Heisenberg uncertainty relation\citep{HUR}, entropic uncertainty relation\citep{EUR}, fine-grained uncertainty relation\citep{FUR}, sum uncertainty relation\citep{SUR} etc, have been employed for Bob's local quantum state. However, when the point of interest is to steer the coherence of the quantum state, uncertainty relations using basis-dependent quantifiers of quantum coherence have to be used. 

Using the $l_1$-norm of quantum coherence\citep{Baumgratz} of a state, $\rho$, i.e. $C^{l_1}(\rho)=\sum_{\substack{i,j\\i\neq j}} |\rho_{ij}|$, the following complementarity relation has been derived in\citep{Mondal},
\begin{equation}
\sum_{i=x,y,z} C_i^{l_1}(\rho) \leq \sqrt{6}
\end{equation}
where the coherence of a single qubit $\rho$ is measured in mutually unbiased basis of dimension $2$, i.e., $\lbrace x.y,z \rbrace$. Now, if Alice and Bob share a bipartite qubit state $\sigma_{AB}$, then after measuring $\Pi_A^a$ on Alice's side, the conditional state on Bob's side becomes,
\begin{align}
\sigma_{B|\Pi_A^a} = &\frac{\operatorname{Tr}_A [(\Pi_A^a \otimes \openone_2) \sigma_{AB}]}{p(\sigma_{B|\Pi_A^a)}} = &\frac{\operatorname{Tr}_A [(\Pi_A^a \otimes \openone_2) \sigma_{AB}]}{\operatorname{Tr} [(\Pi_A^a \otimes \openone_2) \sigma_{AB}]}
\label{cond}
\end{align}
where, $A$ is Alice's choice of measurement and $a$ is the corresponding outcome, and $p(\sigma_{B|\Pi_A^a})$ denotes the probability of getting output $a$ from measurement $A$ on Alice's side. For the joint state, $\sigma_{AB}$ to achieve non-local advantage of quantum coherence using the $l_1$-norm, the sufficient criterion is given by,
\begin{equation}
N^{l_1}= \frac{1}{2} \sum_{i,j,a} p(\sigma_{B|\Pi_{j\neq i}^a}) C_i^{l_1}(\sigma_{B|\Pi_{j\neq i}^a}) \leq \sqrt{6}
\label{Nl1}
\end{equation}
where, $i,j \in \lbrace x,y,z \rbrace$ and $a \in \lbrace 0,1 \rbrace$. Violation of Eq.(\ref{Nl1}) indicates achieving of non-zero NAQC.

Similarly, if quantum coherence of single qubit, $\rho$ is defined by relative entropy\citep{Baumgratz}, $C^E(\rho)=S(\rho_D)-S(\rho)$, where $S(\rho)$ is Von-Neumann entropy of state $\rho$ and $\rho_D$ is the diagonal matrix formed by the diagonal elements of $\rho$ in a fixed basis, then the complementarity relation of coherence in three mutually orthogonal basis, $\lbrace x,y,z \rbrace$ can be constructed as,
\begin{equation}
\sum_{i=x,y,z} C_i^E(\rho) \leq C_2^m \simeq 2.23
\end{equation}
The sufficient condition for the state $\sigma_{AB}$ to achieve NAQC using relative entropy of coherence becomes,
\begin{equation}
N^E= \frac{1}{2} \sum_{i,j,a} p(\sigma_{B|\Pi_{j\neq i}^a}) C_i^E(\sigma_{B|\Pi_{j\neq i}^a}) \leq 2.23
\label{Ne}
\end{equation}
NAQC is attained if the inequality(\ref{Ne}) is violated.

In a similar way, one can define an observable measure of coherence in terms of skew-information of state $\rho$ in a basis of Pauli spinor $\sigma_i$ (i=x,y,z) as\citep{Girolami},
\begin{align*}
C_i^S(\rho)=-\frac{1}{2} \operatorname{Tr}[\sqrt{\rho},\sigma_i]^2=\operatorname{Tr}[\rho.\sigma_i.\sigma_i-\sqrt{\rho}.\sigma_i.\sqrt{\rho}.\sigma_i]
\end{align*}
Hence, complementarity of quantum coherence between different orthogonal basis in $2$ dimensions can be written as,
\begin{equation}
\sum_{i=x,y,z} C_i^S(\rho) \leq 2
\end{equation}
Using the above relation, the criterion for achieving NAQC of the state $\sigma_{AB}$ using skew-information can be given through the violation of the  inequality,
\begin{equation}
N^S= \frac{1}{2} \sum_{i,j,a} p(\sigma_{B|\Pi_{j\neq i}^a}) C_i^S(\sigma_{B|\Pi_{j\neq i}^a}) \leq 2
\label{Ns}
\end{equation}
$\sigma_{AB}$ demonstrates NAQC if it violates any of the three form of inequalities given by Eq.(\ref{Nl1}),(\ref{Ne}),(\ref{Ns}). It can be easily shown that, Alice's choice of measurement is optimal when it is complementary to the choice of measurement for Bob. Achieving NAQC is a task which is asymmetric operationally.

\section{Framework of multiple unsharp measurements at one end} \label{3}

We now describe a scenario introduced in\citep{Shiladitya}, where the no-signalling constraint is no longer applicable. We take the initial state shared between Alice and Bob as $\sigma_{AB}$. Bob holds one particle and the other particle is at Alice's end where several Alices (say, Alice$^1$,Alice$^2$,..etc.) are present. Initially, it is accessed by the first Alice or Alice$^1$. After performing her measurement, Alice$^1$ passes the particle to Alice$^2$ who after performing her measurement, transmits the particle to Alice$^3$, and so on and so forth. Every Alice is independent of any other Alice and they perform unbiased unsharp measurement sequentially to make measurements of all Alice to be equally probable. According to our prescription, only the last Alice performs sharp measurement, because sharp measurement by any other Alice will instantly destroy the entanglement of the pair of particles. Our aim is to find the last Alice upto whom NAQC is achieved between Alice an
 d Bob.

Now we briefly recall the unsharp measurement scheme, which is analogous to weak measurement formalism discussed in\citep{Silva}. The framework of Von-Neumann measurement\citep{Von} implies interaction of the system with the apparatus in a way such that  $|\psi\rangle \equiv \alpha |0\rangle + \beta |1\rangle \rightarrow \alpha |0\rangle \otimes \varphi(q-1) + \beta |1\rangle \otimes \varphi(q+1)$, where $\varphi(q)$ is the pre-measurement state of the apparatus. As per the weak measurement formalism, the reduced state of the system by tracing out apparatus state becomes,
\begin{equation}
\varrho \rightarrow \varrho' = F \varrho + (1-F) (\pi^+ \varrho \pi^+ + \pi^- \varrho \pi^-)
\end{equation}
where $\varrho=|\psi\rangle\langle\psi|$, and the quality factor of the measurement, $F(\varphi)=\int_{-\infty}^{+\infty} \langle \varphi(q+1)|\varphi(q-1) \rangle dq$  implies the extent to which the system remains unaffected post-measurement. The probability of outcomes $\pm$ corresponding to projectors $\pi^{\pm}$ is given by,
\begin{equation}
p_{\pm} = G \langle \psi|\pi^{\pm}|\psi \rangle + \frac{1-G}{2}
\end{equation}
where, $G=\int_{-1}^{+1} \varphi^2 (q) dq$, which quantifies the precision of measurement, i.e., gain of information through measurement. Strong measurement entails $F=0$ and $G=1$. An optimal measurement is that which provides maximum precision for a given quality factor and is captured by the trade-off relation $F^2 + G^2 =1$, as was shown through numerical analysis of various apparatus models
in~\citep{Silva}. 

However, such an optimal trade-off relation emerges naturally within the framework of
 Positive Operator Valued Measurements (POVM) or unsharp measurement formalism~\citep{Shiladitya}. POVM is defined by effect operators $E^i$ which satisfy $0 \leq E^i \leq \openone_2$ and $\sum_i E^i = \openone_2$. In case of dichotomic unsharp measurement,
\begin{equation}
E^{\pm} = \lambda P^{\pm} + \frac{1-\lambda}{2} \openone_2
\label{unsharp}
\end{equation}
where $\lambda$($0< \lambda\leq 1$) is the sharpness parameter and $P^{\pm}$ are projective measurements corresponding to outcomes $\pm$. $\openone_2$ is $2 \times 2$ identity matrix.
According to the non-selective L\"{u}der transformation rule, the state of the system changes as follows:
\begin{equation}
\varrho \rightarrow \varrho' = \sqrt{1-\lambda^2} \varrho + (1-\sqrt{1-\lambda^2}) (P^+ \varrho P^+ + P^- \varrho P^-)
\label{Luder}
\end{equation} 
with probabilities corresponding to $\pm$ outcomes as,
\begin{equation}
p_{\pm} = \operatorname{Tr}[E^{\pm} \varrho] = \lambda ~ \operatorname{Tr}[P^{\pm} \varrho] + \frac{1-\lambda}{2}
\end{equation}
Hence, in comparison to weak formalism, one can find that, $F=\sqrt{1-\lambda^2}$ and $G=\lambda$ satisfy the trade-off relation, $F^2 + G^2 =1$. Sharp measurement implies $\lambda=1$.

\section{Bound on number of Alices exhibiting NAQC} \label{4}

We  consider that Alice$^1$ and Bob share a singlet state  $|\psi^-\rangle = \frac{1}{\sqrt{2}} (|01\rangle - |10\rangle)$ initially, where $\lbrace |0\rangle, |1\rangle \rbrace$ form the eigenbasis of $\sigma_z$. There are say, $n$ number of Alice who can perform unsharp measurements individually, in order to share NAQC through capability of steering quantum coherence of Bob. We assume that except for the $n$-th Alice, the rest  $(n-1)$ of them perform measurements with non-zero unsharpness. Every Alice is independent of the others and is ignorant about the choice of measurement by previous Alices. Therefore, any Alice has to take the average effect of possible settings chosen by previous Alices into account. The possible directions of measurement for each Alice are taken to be $\lbrace x,y,z \rbrace$. The possible choices of basis for Bob in which he determines quantum coherence of his subsystem are purely mutually unbiased basis in $2$-dimensions, i.e., $\lbrace x,y,z \rbrace$ by performing sharp measurements.

\subsection{Sharing of $l_1$-norm of NAQC}

The singlet state (i.e. $\sigma_{A_1 B}=|\psi^-\rangle\langle\psi^-|$) maximally violates Eq.(\ref{Nl1}) and the QM maximum of $N^{l_1}$ is $3$. When Alice$^1$ performs unsharp measurements corresponding to the projectors along $x$, $y$ and $z$ directions, following Eq.(\ref{unsharp}) and using $\lambda=\lambda_1$,  the function at the left hand side of Eq.(\ref{Nl1})  turns out to be
\begin{equation}
N_{A_1 B}^{l_1}= \frac{6 \lambda_1}{1+\lambda_1^2}
\end{equation}
Hence, using the above function, the violation of Eq.(\ref{Nl1}) occurs when $\lambda_1 > \frac{\sqrt{3}-1}{\sqrt{2}}\simeq 0.52$.
The pre-measurement state of Bob and Alice$^2$ depending on the choice of setting for Alice$^1$ i.e. $E_j$, becomes,
\begin{align}
\sigma_{A_1 B} \rightarrow \lbrace \sigma_{A_2 B}^j =& \sqrt{1-\lambda_1^2} ~ \sigma_{A_1 B} + (1- \sqrt{1-\lambda_1^2}) \nonumber\\
& \times [(E^{+}_j \otimes \openone_2) ~ \sigma_{A_1 B} ~ (E^{+}_j \otimes \openone_2) \nonumber\\
&+ (E^{-}_j \otimes \openone_2) ~ \sigma_{A_1 B} ~ (E^{-}_j \otimes \openone_2)] \rbrace_j
\label{pms1}
\end{align}
Here, $E^{\pm}_j$ consists of projectors $P^{\pm}_j$ along the direction of $j\in \lbrace x,y,z \rbrace$.

Now Alice$^2$ performs spin component measurements $E_k$ along direction $k\in\lbrace x,y,z \rbrace$ with unsharpness parameter $\lambda=\lambda_2$. Taking average over all possible choices of Alice$^1$, the NAQC function for Alice$^2$ and Bob turns out to be,
\begin{align}
\overline{N_{A_2 B}^{l_1}} =& \sum_{j} p_j^{A_1} N_{A_2 B}^{l_1} \nonumber\\
=& \frac{1}{2} \sum_{i,j,k,a} p^{A_1}_{j} p(\sigma^{j}_{B|E_{k\neq i}^a}) C_i^{l_1}(\sigma^{j}_{B|E_{k\neq i}^a})
\label{Nl2}
\end{align}
where, $i$ denotes the basis chosen by Bob and the indices $j$ and $k$ represent the directions of measurements chosen by Alice$^1$ and Alice$^2$, respectively. '$a$' is the corresponding outcome of Alice$^2$ whose correlation with Bob is considered here. We take, $i,j,k \in \lbrace x,y,z \rbrace$ and $a \in \lbrace 0,1 \rbrace$. The conditional states,$\lbrace\sigma^{j}_{B|E_{k\neq i}^a}\rbrace_{j=x,y,z}$ can be obtained from $\lbrace \sigma_{A_2 B}^j \rbrace_{j=x,y,z}$ post to the measurement, $E_k$ ($k\in\lbrace x,y,z \rbrace$) by Alice$^2$, using Eq.(\ref{cond}). Since we suppose unbiased input setting for each Alice here,  all possible measurement settings chosen by the previous Alice are equally probable. Thus $p_j^{A_1}=\frac{1}{3} ~ \forall j\in\lbrace x,y,z \rbrace$. From Eq.(\ref{Nl2}), we get $\overline{N_{A_2 B}^{l_1}}=\frac{2 \lambda_2 (1+ 2\sqrt{1- \lambda_1^2})}{1+ \lambda_2^2}$.

If we consider the effect of NAQC upto Alice$^2$ in the given scenario,  we have to choose sharp measurement for Alice$^2$, i.e., $\lambda_2=1$. We observe that $\overline{N_{A_2 B}^{l_1}} > \sqrt{6}$ when $\lambda_1 < \frac{1}{2} \sqrt{2\sqrt{6}-3} \simeq 0.69$. So, in order to achieve NAQC by Alice$^1$ and Alice$^2$, or in other words to steer the $l_1$-norm of coherence of Bob by both the Alices, we obtain a region for $\lambda_1$, i.e., $\lambda_1\in (0.52,0.69)$. It is also seen from here that whenever $\lambda_1 > 0.52$, the 2nd Alice attains NAQC for $\lambda_2 > 0.63$. Hence, in this range, the QM maximum of $\overline{N_{A_2 B}^{l_1}}$ is $2.71$.

Now we introduce the 3rd Alice or Alice$^3$ in this context where Alice$^1$ and Alice$^2$  perform unsharp measurements. In this case, the pre-measurement state of Alice$^3$ based on settings of Alice$^1$ and Alice$^2$, i.e., $E_j$ and $E_k$ respectively, can be written as
\begin{align}
\lbrace \sigma_{A_2 B}^j \rbrace_{j} \rightarrow \lbrace \sigma_{A_3 B}^{jk} =& \sqrt{1-\lambda_2^2} ~ \sigma_{A_2 B}^{j} + (1- \sqrt{1-\lambda_2^2}) \nonumber\\
& \times [(E^{+}_{k} \otimes \openone_2) ~ \sigma_{A_2 B}^{j} ~ (E^{+}_{k} \otimes \openone_2) \nonumber\\
&+ (E^{-}_{k} \otimes \openone_2) ~ \sigma_{A_2 B}^{j} ~ (E^{-}_{k} \otimes \openone_2)] \rbrace_{j,k}
\end{align}
Here, $E_k^{\pm}$ ($k\in\lbrace x,y,z \rbrace$) are the effect operators corresponding to projectors, $P_k^{\pm}$ with outcomes $\pm$1.

Alice$^3$ performs measurements along $l\in\lbrace x,y,z \rbrace$ components of spin with precision $\lambda=\lambda_3$. The average effect of Alice$^1$ and Alice$^2$ on the NAQC between Alice$^3$ and Bob turns out to be
\begin{align}
\overline{N_{A_3 B}^{l_1}} =& \sum_{j,k} p_j^{A_1} p_k^{A_2} N_{A_3 B}^{l_1} \nonumber\\
=& \frac{1}{2} \sum_{i,j,k,l,a} p^{A_1}_{j} p^{A_2}_{k} p(\sigma^{jk}_{B|E_{l\neq i}^a}) C_i^{l_1}(\sigma^{jk}_{B|E_{l\neq i}^a})
\label{Nl3}
\end{align}
where, $i$ represents Bob's choice of basis, the indices $j$,$k$,$l$ are settings corresponding to Alice$^1$, Alice$^2$ and Alice$^3$ respectively, and $a$ denotes the outcome of measurement $E_l$ done by the 3rd Alice. The correlation between Alice$^3$ and Bob is of our interest. We consider, $i,j,k,l\in\lbrace x,y,z \rbrace$ and $a\in\lbrace 0,1 \rbrace$. We get the conditional states $\lbrace\sigma^{jk}_{B|E_{l\neq i}^a}\rbrace_{j,k}$ from $\lbrace \sigma_{A_3 B}^{jk} \rbrace_{j,k}$ after the measurement $E_l$ by Alice$^3$, using Eq.(\ref{cond}). Unbiased measurement settings for every Alice leads us to assume $p_j^{A_1}=p_k^{A_2}=\frac{1}{3} ~ \forall j,k\in\lbrace x,y,z \rbrace$ in the $3$-measurement scenario. From Eq.(\ref{Nl3}), we obtain $\overline{N_{A_3 B}^{l_1}}= \frac{2 \lambda_3 [1+ 2\sqrt{1- \lambda_1^2} + 2\sqrt{1- \lambda_2^2} + 4 \sqrt{(1- \lambda_1^2) (1- \lambda_2^2)}]}{3 (1+ \lambda_3^2)}$. It follows from the above expression that when Alice$^1$ and Alice$^2$ get the effect of NAQC through the violation of Eq.(\ref{Nl1}), the NAQC value between Alice$^3$ and Bob can not be greater than $\sqrt{6}$. In fact, when Alice$^1$ and Alice$^2$ have the least amount of NAQC, the QM maximum of $\overline{N_{A_3 B}^{l_1}}$ turns out to be $2.30$. Therefore, it is impossible for Alice$^3$ to steer the coherence of Bob when Alice$^1$ and Alice$^2$ are able to do it.

It can be easily checked that if one of the restrictions imposed by precisions $\lambda_1$ and $\lambda_2$ is relaxed, then Alice$^3$ and Bob can reveal NAQC. Hence, any two of the Alices can demonstrate NAQC while the other can not. Bob's coherence can be steered by any of the following pair of Alices: (Alice$^1$, Alice$^2$), (Alice$^2$, Alice$^3$), (Alice$^1$, Alice$^3$). In other words, not more than two observers (Alices) can show the feature of NAQC on Bob when the $l_1$-norm of coherence measure is probed in the given framework.

\subsection{Sharing of relative entropy of NAQC}

We now focus on sharing of NAQC using the relative entropy as coherence measure  demonstrated by the violation of Eq.(\ref{Ne}). We ask whether the bound on the number of Alices who can independently steer the relative entropy of coherence of Bob in the foregoing scenario is similar as that of the previous case utilizing the $l_1$-norm of coherence. Keeping the possible choices of measurement settings for all Alices and the single Bob unchanged, the singlet state ($\sigma_{A_1 B}=|\psi^-\rangle\langle\psi^-|$) gives maximum QM violation of Eq.(\ref{Ne}) i.e. $N^E (\sigma_{A_1 B})=3$.

When Alice$^1$ performs unsharp measurement $E_j$ along direction $j\in\lbrace x,y,z \rbrace$ satisfying Eq.(\ref{unsharp}) with $\lambda=\lambda_1$, then the NAQC function at the left hand side of Eq.(\ref{Ne}) becomes,
\begin{equation}
N_{A_1 B}^E= log_2 (e) ~ [\frac{12 \lambda_1}{1+\lambda_1^2} tanh^{-1}(\lambda_1)-6 tanh^{-1}(\lambda_1^2)]
\end{equation}
Eq.(\ref{Ne}) is violated when $\lambda_1>0.65$.

Now Alice$^2$ measures $E_k$ ($k\in\lbrace x,y,z \rbrace$) with precision $\lambda=\lambda_2$ on the pre-measurement states given by Eq.(\ref{pms1}). The average NAQC function between Alice$^2$ and Bob over all possible settings of Alice$^1$ becomes
\begin{align}
\overline{N_{A_2 B}^{E}} =& \sum_{j} p_j^{A_1} N_{A_2 B}^{E} \nonumber\\
=& \frac{1}{2} \sum_{i,j,k,a} p^{A_1}_{j} p(\sigma^{j}_{B|E_{k\neq i}^a}) C_i^{E}(\sigma^{j}_{B|E_{k\neq i}^a})
\label{Ne1}
\end{align}
where $i\in\lbrace x,y,z \rbrace$ is the basis chosen by Bob in which he measures the relative entropy of coherence on the conditional state given by $\sigma^{j}_{B|E_{k\neq i}^a} \forall j$. Alice$^1$ and Alice$^2$ measure along $j,k\in\lbrace x,y,z \rbrace$, respectively. The result of the measurement done by Alice$^2$ is $a\in\lbrace 0,1 \rbrace$. As mentioned before, unbiased or equiprobable measurement settings for Alice$^1$ allows us to consider $p_j^{A_1}=\frac{1}{3} ~ \forall j\in\lbrace x,y,z \rbrace$.

We obtain from Eq.(\ref{Ne1}) that,
\begin{align}
\overline{N_{A_2 B}^{E}}=& 2[1 + log_2 (e) ~ (\frac{2 \lambda_2}{1+\lambda_2^2} tanh^{-1}(\lambda_2)- tanh^{-1}(\lambda_2^2)) \nonumber\\
&+ (\frac{1}{2}-\frac{\sqrt{1-\lambda_1^2} \lambda_2}{1+\lambda_2^2}) ~ log_2 (\frac{1}{2}-\frac{\sqrt{1-\lambda_1^2} \lambda_2}{1+\lambda_2^2}) \nonumber\\
&+ (\frac{1}{2}+\frac{\sqrt{1-\lambda_1^2} \lambda_2}{1+\lambda_2^2}) ~ log_2 (\frac{1}{2}+\frac{\sqrt{1-\lambda_1^2} \lambda_2}{1+\lambda_2^2})]
\end{align}

It can be checked that, Alice$^2$ and Bob can not achieve NAQC in the range $\lambda_1>0.65 ~ \forall \lambda_2$. The QM maximum of $\overline{N_{A_2 B}^{E}}$ is $1.94$ which is much smaller than the upper bound of Eq.(\ref{Ne}), i.e., $2.23$, and is attained in the  scenario where Alice$^1$ and Bob have negligible violation of Eq.(\ref{Ne}) and Alice$^2$ measures as sharp as possible. Hence only one Alice can share NAQC with Bob when relative entropy is employed as a coherence measure in the framework of multiple POVMs. Here we observe that the bound on the number of Alices is different from the earlier case using the $l_1$-norm as measure of coherence.

\subsection{Sharing of skew-information of NAQC}

Here we investigate the sharing of NAQC applying skew-information as quantifier of quantum coherence. The correlation can be viewed through the QM violation of Eq.(\ref{Ns}). We enquire in the given scenario as to how many Alices can independently exhibit NAQC with a single Bob. The possible choices of settings for Alices and Bob are the kept same as before. The singlet state, $\sigma_{A_1 B}$ (i.e. $|\psi^-\rangle\langle\psi^-|$) provides the maximal QM violation of Eq.(\ref{Ns}), i.e., $N_{A_1 B}^S =3$.

If Alice$^1$ chooses to measure along $j\in\lbrace x,y,z \rbrace$  with unsharpness $\lambda=\lambda_1$,  the left hand side of the inequality(\ref{Ns}) turns out to be
\begin{equation}
N_{A_1 B}^S=\frac{6 \lambda_1^2}{1+\lambda_1^2}
\end{equation}
which violates the upper bound of Eq.(\ref{Ns}) when $\lambda_1 > \frac{1}{\sqrt{2}} \simeq 0.71$.

Now, consider that Alice$^2$ measures along $k\in\lbrace x,y,z \rbrace$ with unsharpness parameter $\lambda=\lambda_2$ considering  the bipartite states given by Eq.(\ref{pms1}). As Alice$^2$ has no knowledge about the choice of setting of the previous Alice, the correlation function of NAQC between Alice$^2$ and Bob is taken by averaging over all possible choices by Alice$^1$, and can be expressed as,
\begin{align}
\overline{N_{A_2 B}^{S}} =& \sum_{j} p_j^{A_1} N_{A_2 B}^{S} \nonumber\\
=& \frac{1}{2} \sum_{i,j,k,a} p^{A_1}_{j} p(\sigma^{j}_{B|E_{k\neq i}^a}) C_i^{S}(\sigma^{j}_{B|E_{k\neq i}^a})
\label{Ns1}
\end{align}
where, $i,j,k\in\lbrace x,y,z \rbrace$ denote  the possible choices of basis by Bob, Alice$^1$ and Alice$^2$, respectively. $a\in\lbrace 0,1 \rbrace$ is the output corresponding to dichotomic input of Alice$^2$. Unbiased input setting dictates that the probability of choosing any one (i.e., $j\in\lbrace x,y,z \rbrace$) out of three settings by Alice$^1$, i.e., $p^{A_1}_{j}$ is equal for all possible choices. So, as considered by Alice$^2$, $p_j^{A_1}=\frac{1}{3} ~ \forall j$.

From Eq.(\ref{Ns1}), we get
\begin{equation}
\overline{N_{A_2 B}^{S}} = \frac{2[1+2\lambda_2^2-\sqrt{(1-\lambda_2^2)^2 +4 \lambda_1^2 \lambda_2^2}]}{1+\lambda_2^2}
\end{equation}

We see that the pair of Alice$^2$ and Bob is unable to reveal NAQC when Alice$^1$ shows the correlation of NAQC with Bob, i.e., in the range of $\lambda_1 \in (0.71,1)$. The QM maximum of $\overline{N_{A_2 B}^{S}}$ is $1.59$ occurring  when Alice$^1$ and Bob marginally violate Eq.(\ref{Ns}) and Alice$^2$ performs sharp dichotomic qubit measurements.
Hence, not more than one Alice can steer the skew-information of Bob in the given scenario. This result is similar to that of relative entropy of NAQC.

%%%%%%%%%%%%%%%%%%%%%%%%%%%%%%%%%%%%%%%%%%%%%%%%%%%%%%%%%%%%%%%%%%%%%%%%%%%%%%%%%%%%%%%%%%%%%
%%%%%%%%%%%%%%%%%%%%%%%%%%%%%  up to here  %%%%%%%%%%%%%%%%%%%%%%%%%%%%%%%%%%%%%%%%%
%%%%%%%%%%%%%%%%%%%%%%%%%%%%%%%%%%%%%%%%%%%%%%%%%%%%%%%%%%%%%%%%%%%%%%%%%%%%%%%%%%%%%%%%%%%

\section{Conclusions} \label{5}

Sequential sharing of nonlocal quantum correlations could be of relevance in practical information theoretic protocols involving secret key generation among multiple parties, and randomness certification~\citep{random,random1}. In the present work, we investigate the sequential sharing of nonlocal advantage of quantum coherence (NAQC)~\citep{Mondal} which is a correlation stronger than Bell nonlocality~\citep{Hu, Hu1}. We consider a bipartite entangled state in the $\mathbb{C}^2 \otimes \mathbb{C}^2$ Hilbert space where multiple Alices perform sequential POVMs on half of the entangled pair, and a single Bob measures coherence on the other half in a particular basis.  We show that at most two Alices can demonstrate NAQC with Bob using the $l_1$-norm as measure of coherence. We further show that not more than one Alice is able to share either the relative entropy of NAQC or the skew-information of NAQC with Bob in the same configuration. 

Our results indicate a qualitative difference between the $l_1$-norm of coherence with the other two measures, i.e., the relative entropy and skew information, when coherence is employed as a resource for generating nonlocal advantage in information processing tasks. It is known from earlier results that Bell-CHSH non-locality can be shared between at most two observers at one end~\citep{Silva,Shiladitya,Expt1,Expt2}, whereas quantum steering has been conjectured to be demonstrated with at most $n$ number of observers at one end when a steering inequality with $n$-measurement settings per party is employed~\citep{Das}. Based on these results it might be expected that the bound on the number of observers is inherently connected to the strength of the nonlocal correlation, with increase in the strength of the correlation causing the number of observers being able to share it to be decreased. Our present analysis supports such an intuition to the extent that the relative entropy and skew information of NAQC can only be shared by one Alice with a single Bob on the other side. Further work, perhaps also involving higher dimensional quantum states, is needed to shed more light on this absorbing issue of sharing of quantum correlations among multiple observers without being constrained by the no-signalling condition.

{\it Acknowledgements:} SD is thankful to  Shiladitya Mal for his fruitful suggestions and acknowledges financial support through INSPIRE Fellowship from Department of Science and Technology, Govt. of India (Grant No. C/5576/IFD/2015-16). ASM acknowledges Project No.: DST/ICPS/Qust/2018/98 from Department of Science and Technology, Government of India.

\bibliography{NAQC}

\end{document}